# Double-Rashba materials for nanocrystals with bright ground-state excitons

Michael W. Swift[1]*, Peter C. Sercel[2]*, Alexander L. Efros[1], John L. Lyons[1], and David J. Norris[3]*

[1]Center for Computational Materials Science, U.S. Naval Research Laboratory, Washington, D.C. 20375 USA.
[2]Center for Hybrid Organic–Inorganic Semiconductors for Energy, Golden, Colorado 80401, USA.
[3]Optical Materials Engineering Laboratory, Department of Mechanical and Process Engineering, ETH Zurich, 8092 Zurich, Switzerland.

*email: michael.swift@nrl.navy.mil; pcsercel@gmail.com; dnorris@ethz.ch

**ABSTRACT**

While nanoscale semiconductor crystallites provide versatile fluorescent materials for light-emitting devices, such nanocrystals suffer from the 'dark exciton'—an optically inactive electronic state into which the nanocrystal relaxes before emitting. Recently, a theoretical mechanism was discovered that can potentially defeat the dark exciton. The Rashba effect can invert the order of the lowest-lying levels, creating a bright excitonic ground state. To identify materials that exhibit this behavior, here we perform an extensive high-throughput computational search of two large open-source materials databases. Based on a detailed understanding of the Rashba mechanism, we define proxy criteria and screen over 500,000 solids, generating 173 potential 'bright-exciton' materials. We then refine this list with higher-level first-principles calculations to obtain 28 candidates. To confirm the potential of these compounds, we select five and develop detailed effective-mass models to determine the nature of their lowest-energy excitonic state. We find that four of the five solids (BiTeCl, BiTeI, $Ga_2Te_3$, and $KIO_3$) can yield bright ground-state excitons. Our approach thus reveals promising materials for future experimental investigation of bright-exciton nanocrystals.

Semiconductors emit light when electrons find holes and recombine. This occurs more readily if each electron is Coulombically bound to a hole, forming an exciton. By confining this exciton in a nanoscale structure, radiative recombination can be further enhanced. This has motivated the development of semiconductor nanocrystals (or colloidal quantum dots)[1-3], which now provide versatile emitters for light-emitting diodes[4], lasers[5,6], and quantum technology[7]. However, nanocrystals suffer from a fundamental problem: the dark exciton[8]. The electron–hole exchange interaction splits the exciton into a series of sublevels known as fine structure. Because radiative recombination from the lowest-energy sublevel is dipole forbidden, emission from this 'dark' ground state is slow and inefficient. In bulk semiconductors, 'bright' sublevels are only marginally higher in energy, and hence, thermally accessible, limiting the impact of the dark exciton. In nanostructures, exciton confinement increases the fine-structure splitting, concentrating the excitation in the dark state and reducing nanocrystal brightness[9].

To avoid this problem, experimental and theoretical efforts have sought materials or mechanisms (external fields[10], strain[11], dopants[12], *etc.*) to brighten the dark state. A bright ground state could also be achieved by flipping the order of the bright and dark states, but extensive efforts failed to reveal such a level inversion, cultivating the belief that the ground-state exciton is always dark in nanocrystals.

However, this conclusion has recently been reexamined due to the observation of exceptional brightness in cesium lead halide perovskite nanocrystals[13-17]. A bright ground-state exciton was proposed as the explanation, and theoretical efforts discovered a potential mechanism—the Rashba effect—for bright–dark inversion[15]. While its actual role in perovskite nanocrystals has been questioned[18-20], further theoretical modeling has confirmed that the Rashba effect can produce bright ground-state excitons under certain conditions[21-24]. Thus, a general route to brighter semiconductor nanocrystals is now known. The challenge is to identify materials with the proper characteristics.

Here we perform an extensive search for such materials by exploiting two open-source databases, the Materials Project[25] and Aflowlib[26,27], which together contain properties for over 500,000 inorganic compounds. By screening these databases against a series of criteria for the Rashba effect and efficient optical emission, we identify 173 promising targets. High-level first-principles theory further refines this list to 28 bright-exciton candidates. To confirm the potential of these compounds, we select five and develop detailed effective-mass models to determine the nature of the ground-state exciton in nanocrystals. We find that four of the five compounds (BiTeCl, BiTeI, $Ga_2Te_3$, and $KIO_3$) can lead to bright ground-state excitons. Our approach thus identifies materials for future experimental investigation of bright-exciton nanomaterials.

**Requirements for bright ground-state excitons**

If the electron–hole exchange interaction is the only influence on the exciton fine structure, a dark ground state results. For the Rashba effect to invert the level order, strong spin–orbit coupling and inversion asymmetry are required. Charge carriers moving in such a material can feel an effective magnetic field, causing spin splittings in the electronic states[28]. This effect must occur both for electron and hole motion in the conduction and valence bands, respectively, which we call the 'double-Rashba' requirement. As the electron and hole are bound together as an exciton, different exciton motions can be considered. Prior predictions of bright ground-state excitons relied on the electron–hole center-of-mass motion[15]. However, our modeling indicates that bright excitons can also arise due to the relative electron–hole motion (Supplementary Section S1). Interestingly, these two Rashba mechanisms act oppositely on the exciton fine structure, allowing the bright exciton to be further optimized by nanostructure size and shape.

Specifically, the center-of-mass motion can cause level inversion in nanostructures with sizes larger than the exciton Bohr radius. This case (which we label type I) was previously proposed to explain the exceptional brightness of 14 nm cube-shaped $CsPbBr_3$ nanocrystals[15,21,22]. In contrast, the relative electron–hole motion can invert the levels for excitons under quasi-two-dimensional (quasi-2D) weak confinement (type II). For example, in quasi-2D nanostructures (such as quantum wells or nanoplatelets), the exciton is weakly confined within a plane but strongly confined out of plane. If the exciton has negligible center-of-mass motion, we find (Supplementary Section S1) that the lowest sublevel can be bright, with radiative recombination that is both momentum and spin allowed[24]. In such quasi-2D shapes, excitons can also exhibit giant oscillator strength[29], leading to very fast, bright emission.

## Screening methodology

The Materials Project and Aflowlib databases provide powerful tools for materials discovery. However, spin–orbit coupling, required for the Rashba effect, is not included. Thus, based on our knowledge of the inversion mechanisms, we defined proxy criteria to screen for materials likely to satisfy the double-Rashba requirements.

Figure 1 indicates the number of compounds remaining in the databases after each criterion is applied. The procedure for the Materials Project database is described here; the process for Aflowlib is similar (see Supplementary Section S2 and Tables S2–S4). *Step 1*: we selected all insulators and semiconductors. Although prior work focused on topological insulators to identify materials with 'giant' Rashba effects[30], we were less restrictive as we need only modest splittings. *Step 2*: we screened for compounds containing heavy *p*-block elements. While any heavy atom would provide strong spin–orbit coupling, materials with *p*-block elements are more likely to give rise to low effective masses and symmetry-allowed band-edge transitions. *Step 3*: we kept only polar materials. For a bulk solid (without surface or structural inversion asymmetry) to exhibit the Rashba effect, it should have polar sites with inversion asymmetry, represented by the polar space groups (Supplementary Table

S3). Notably, this eliminated the inorganic halide perovskites like CsPbBr$_3$ due to their centrosymmetry (although recent theoretical results suggest polar phases could form in a nanocrystal under tensile strain[31]). *Step 4*: we focused on direct-bandgap solids to ensure that exciton recombination is dipole allowed. In practice, this also excluded materials with unknown band structures. *Step 5*: we analyzed the atom-projected density of states at the valence-band maximum and conduction-band minimum to check that Rashba splittings occur at these band edges, thus contributing to the exciton fine structure.

## Density-functional-theory refinement

Our screening procedure produced 173 targets from the two databases. First-principles calculations (Methods and Supplementary Section S3) were then performed to confirm double-Rashba behavior. We first applied density functional theory (DFT) using the semi-local Perdew-Burke-Ernzerhof (PBE) functional[32] with spin–orbit coupling (PBE+SOC). If the double-Rashba criterion remained satisfied, further checks with the hybrid Heyd-Scuseria-Ernzerhof (HSE) functional[33] with spin–obit coupling (HSE+SOC) were completed. This process yielded 26 candidates. Because BiTeCl, BiTeBr, and KSnAs were included, we added BiTeI and KPbAs to our analysis and found two more double-Rashba solids. Thus, in total, 28 potential bright-exciton materials were identified (Table 1). Only the hybrid organic–inorganic halide perovskites and the bismuth tellurohalides were previously known as double-Rashba materials.

## Fine-structure determination

The exciton fine structure for a given nanocrystal can then be calculated with effective-mass models if the bulk band parameters are known. Thus, for each identified candidate, we determined 16 material parameters using HSE+SOC (Supplementary Tables S5–S7). Table 1 presents five parameters particularly relevant for bright-exciton materials: the bandgap energy, $E_g$, which relates to the emission energy; the exciton Rashba energy, $\varepsilon_R$, which quantifies the Rashba-interaction strength (see Supplementary Section S3); the Kane energy, $E_\mathrm{p}$, which sets the emission rate; the exciton Bohr radius, $a$, which defines the confinement regime; and the spin-texture helicity, STH, which describes whether the Rashba spin textures in the conduction and valence bands are co-helical (STH$^+$) or contra-helical (STH$^-$). Our work suggests that STH$^-$ and STH$^+$ lead to type-I and type-II behavior, respectively (Supplementary Sections S1 and S5). Thus, materials with STH$^-$ can exhibit bright ground-state excitons under intermediate confinement, and those with STH$^+$ require quasi-2D nanostructures.

To predict the excitonic level order, effective-mass models already exist for well-studied nanomaterials (such as CdSe and $CsPbBr_3$) and can be adapted for candidates in the same space group. For other crystal symmetries, models must be constructed. Here, we focus on five candidates from Table 1: BiTeBr, BiTeCl, BiTeI, $Ga_2Te_3$, and $KIO_3$.

We begin with the bismuth tellurohalides, as they exhibit a strong Rashba effect[34] and effective-mass models have already been developed to describe the spin physics of electrons and holes[35,36]. We also confirmed that the BiTeX compounds share similar crystal structures (Fig. 2a shows BiTeI) and display double-Rashba behavior (revealed by the spin-polarized 'double-well' shapes at both the conduction and valence band edges in their band diagrams; see Fig. 2b for BiTeI). Thus, by developing a model for BiTeX nanocrystals, we can potentially predict three bright-exciton nanocrystals.

Due to the strong Rashba effect in nanoscale BiTeX, we exploited a non-perturbative description of Rashba excitons[24]. Supplementary Section S4 describes our calculations for the exciton binding energies, exciton fine structures (short- and long-range exchange interactions and the Rashba interaction), and bright-exciton properties (oscillator strengths and radiative lifetimes). A Python implementation of the model is also provided. As expected for materials with co-helical spin textures (Table 1), we find that the relative exciton motion in BiTeX lowers the energy of the bright exciton. Level inversion can then occur *via* a type-II mechanism, suggesting quasi-2D nanostructures should be targeted[24].

Therefore, we considered monolayer-thick BiTeX disks of radius $R$. For such thin nanocrystals, we used DFT (instead of the values in Table 1) to obtain bandgaps, effective masses, and Rashba coefficients. These parameters (Supplementary Table S10) were then inserted into our exciton model to predict the fine structure. Figure 2c–d shows that the level energies oscillate for BiTeX with increasing disk size. Moreover, the bright exciton becomes the ground state at specific disk radii for BiTeCl and BiTeI. These oscillations occur as radial nodes are added to the excitonic wavefunctions with increasing disk size. Such nodes appear at different radii for the bright and dark states. After a node is added, the wavefunction 'fits' poorly in the disk, raising its confinement energy until the disk increases in size (Supplementary Section S4). In BiTeBr, these oscillations are insufficient to invert the level order.

These results clearly demonstrate that our process can identify bright-exciton materials (BiTeCl and BiTeI). However, despite the large $\varepsilon_R$ for BiTeX (Table 1), the size of the bright–dark splitting is small (<0.4 meV). To seek larger splittings, which can enhance nanocrystal brightness, we considered $Ga_2Te_3$. Its electron–hole exchange interaction is about five times smaller than in BiTeX, allowing the Rashba effect to invert the level order more easily.

Figure 3a shows the crystal structure of bulk $Ga_2Te_3$[37,38]. Its excitonic and photoluminescence properties have not previously been reported. Our calculated band structure confirms double-Rashba behavior (Fig. 3b). To determine the fine structure in $Ga_2Te_3$ nanocrystals, the Rashba term can be included perturbatively, as $\varepsilon_R$ is modest (Table 1). Generalizing an approach developed for perovskite nanocrystals[21,22], we built an exciton model for monoclinic $Ga_2Te_3$. Due to its contra-helical spin texture (Table 1), we targeted a type-I mechanism. We considered cube-shaped $Ga_2Te_3$ nanocrystals under intermediate confinement and calculated the exciton fine structure as a function of the nanocrystal side length, $L$ (ref. 21). Figure 3c plots the fine-structure energy corrections of the three bright sublevels (labelled $YZ^L$, $X$, and $YZ^U$ according to the orientation of their transition dipoles) relative to the dark exciton in bulk Ga$_2$Te$_3$ (dashed line). Due to the low crystallographic symmetry, the states $YZ^U$ and $YZ^L$ possess non-orthogonal transition dipoles while the dark state mixes with the $X$ state, leading to a size-dependent shift of the dark state and an avoided crossing, even within the assumption of a size-independent exciton Rashba energy. (For details, including dark-state activation and bright state mixing by the monoclinic symmetry and long-range exchange for bright states with non-orthogonal transition dipoles, see Supplementary Section S6.) More importantly, with increasing size, all three bright sublevels shift below the dark state. For $L > 3.8$ nm, the bright level with the largest oscillator strength ($YZ^L$) is the ground state. Its radiative lifetime decreases with increasing nanocrystal size, dropping below 3 ns at $L = 18$ nm (Fig. 3d). These results predict a bright–dark splitting (~4 meV) in Ga$_2$Te$_3$ that is an order of magnitude larger than in BiTeCl and BiTeI.

Finally, we investigated potassium iodate (KIO$_3$, Fig. 4a). KIO$_3$ is the candidate in Table 1 with the largest Kane energy (other than the halide perovskites), so its radiative recombination should be very fast. However, one might also expect this ionic compound to exhibit flat electronic bands due to localization of the electronic states on K$^+$ and IO$_3^-$ ions. Instead, the calculated band structure (Fig. 4b) reveals moderate electron and hole effective masses ($m_e = 0.58\ m_0$ and $m_h = 0.87\ m_0$, respectively, where $m_0$ is the free-electron mass) and double-Rashba behavior. The crystal consists of corner-sharing IO$_6$ octahedra with three short and three long I–O bonds. These octahedra form a network that allows delocalized electronic states, and their asymmetry provides a polar environment for the Rashba effect. To determine the fine structure, we employed a recent quasi-cubic perovskite model[22] and targeted a type-I mechanism (KIO$_3$ is STH$^-$; Table 1). The symmetry of KIO$_3$ forbids mixing between the dark and bright states, so the only energy shift of the dark state is due to the exciton Rashba energy, which we have assumed to be a constant independent of nanocrystal size. For cube-shaped nanocrystals of $KIO_3$ under intermediate confinement[21] (Supplementary Section S7), we predict a bright ground state for $L > 1.9$ nm, with bright–dark splittings of ~10 meV and radiative lifetimes below 40 picoseconds in large nanocrystals.

**Discussion**

For four of the five candidates treated with effective-mass models, we found nanocrystals that exhibit bright ground-state excitons. This confirms that Table 1 presents a rich resource for identification of bright-exciton materials. The presence of $(CH_3NH_3)PbI_3$ and $(CH(NH_2)_2)PbI_3$, both exhibiting $STH^-$, suggests that cube-shaped nanocrystals of such hybrid perovskites could exhibit bright ground states. Furthermore, the appearance of $(CH_3NH_3)SnI_3$ with $STH^+$ indicates that emerging 2D tin-based halide perovskites, such as $(PEA)_2SnI_4$ (ref. 39), should be explored for bright lead-free emitters. Many candidates may also be extended into families through rational chemical substitution, as we demonstrated with BiTeI and KPbAs.

The four bright-exciton materials predicted require experimental confirmation. BiTeCl and BiTeI present examples of a previously unidentified class of type-II bright ground-state emitters. $Ga_2Te_3$ and $KIO_3$ offer fast radiative recombination at infrared and ultraviolet wavelengths, respectively.

Our list of candidates is far from complete. Because we relied on conservative proxies for double-Rashba behavior, additional bright-exciton materials were likely missed. Such open-source databases are also constantly expanding, increasing the search pool. Nevertheless, the success of our search procedure despite the lack of spin–orbit-coupling information in the databases illustrates the power of combining physical understanding with high-throughput techniques.

## Online content

Methods, additional references, source data, extended data, supplementary information, peer-review information, and statements of data and code availability are presented in the online version of this paper.

**Methods**

All first-principles calculations used the Vienna *ab-initio* simulation package (VASP)[40] with projector-augmented wave pseudopotentials (PAW)[41]. Semi-local calculations used the PBE functional[32], while the hybrid functional calculations used the HSE functional[33] with the standard 25% mixing and 0.2 Å$^{-1}$ screening. The plane-wave cutoff was 520 eV, and spin–orbit coupling was included using the default VASP spinor-based methodology. Brillouin-zone sampling was automatically generated by the open-source library pymatgen with a reciprocal density of 64 *k*-points per Å$^{-3}$ of the reciprocal lattice. In some cases doubling the *k*-point sampling density was required for accurate relaxation in the PBE+SOC and HSE+SOC steps; $(CH_3NH_3)PbI_3$ and $Rb_3Bi_7Pb_3I_{10}O_{10}$ from Table 1 used a doubled *k*-mesh.

The electronic convergence threshold was $10^{-5}$ eV, and the force convergence threshold for relaxation was 0.01 eV/Å. All other parameters were set using Materials Project parameters as described by the Materials Project input sets in pymatgen.io.vasp.sets[42].

## Method references

## Data availability

All data that support the findings of this study are available from the corresponding authors upon request.

## Acknowledgements

This work was supported by the US Office of Naval Research (ONR) through the Naval Research Laboratory's Basic Research Program, by the Swiss National Science Foundation under Award Number 200021-188593, and by the Center for Hybrid Organic–Inorganic Semiconductors for Energy (CHOISE), an Energy Frontier Research Center funded by the Office of Basic Energy Sciences, Office of Science within the US Department of Energy, which supported the development of the analytical short-range exchange model for monoclinic $Ga_2Te_3$ and partly supported the development of the analytical fine-structure models generally. M.W.S, Al.L.E. and J.L.L. also acknowledge support from the Laboratory–University Collaboration Initiative (LUCI) program of the Department of Defense (DoD) Basic Research Office. Computations were performed at the DoD high-performance computing centers. We thank Dr. Noam Bernstein for technical assistance and discussions.

**Author contributions**

D.J.N. and Al.L.E. conceived the project. M.W.S. designed and conducted the high-throughput search, developed the models, and performed the calculations with input from P.C.S., J.L.L. and Al.L.E. M.W.S. and D.J.N. wrote the manuscript. All authors contributed to the discussion of the results and to the revision of the manuscript.

**Additional information**

Supplementary information is available in the online version of the paper. Reprints and permissions information is available online at www.nature.com/reprints. Correspondence and requests for materials should be addressed to M.W.S. or D.J.N.

**Competing financial interests**

The authors declare no competing financial interests.

# Tables and figures

| Compound | $E_g$ (eV) | $\varepsilon_R$ (meV) | $E_p^{max}$ (eV) | $a$ (nm) | STH |
|---|---|---|---|---|---|
| BiTeCl | 1.15 | 54.4 | 7.36 | 5.47 | + |
| BiTeBr | 0.97 | 73.0 | 6.02 | 6.26 | + |
| BiTeI | 0.85 | 112 | 4.65 | 7.31 | + |
| CsSiTe$_3$ | 1.98 | 6.11 | 0.10 | 0.42 | − |
| Rb$_3$Bi$_7$Pb$_3$I$_{10}$O$_{10}$ | 1.60 | 61.8 | 0.36 | 0.17 | + |
| KInI$_4$ | 3.33 | 0.13 | 7.22 | 0.17 | − |
| CaPbI$_6$ | 2.41 | 10.2 | 0.26 | 0.41 | + |
| (CH$_3$NH$_3$)PbI$_3$ | 1.37 | 6.21 | 12.6 | 2.23 | − |
| (CH$_3$NH$_3$)SnI$_3$ | 1.13 | 3.33 | 12.1 | 3.08 | + |
| Sn$_4$SF$_6$ | 3.30 | 0.82 | 2.45 | 0.40 | − |
| KSnAs | 0.79 | 7.87 | 1.04 | 5.67 | + |
| NaSn$_4$(PO$_4$)$_3$ | 4.82 | 2.33 | 0.53 | 0.17 | − |
| LiGa$_3$Te$_5$ | 1.67 | 0.48 | 2.26 | 3.29 | − |
| CdGa$_6$Te$_{10}$ | 1.25 | 0.65 | 1.92 | 2.27 | − |
| Ga$_2$Te$_3$ | 1.16 | 1.75 | 4.18 | 2.81 | − |
| RbCdI$_3$(H$_2$O) | 3.50 | 0.09 | 11.0 | 0.46 | − |
| CaGa$_6$Te$_{10}$ | 1.58 | 0.06 | 4.62 | 1.70 | + |
| Ga$_2$HgTe$_4$ | 0.86 | 1.38 | 0.98 | 2.18 | + |
| Li$_5$SbS$_4$ | 1.96 | 1.10 | 10.3 | 1.01 | + |
| HgI$_3$(NH$_4$)(H$_2$O) | 2.44 | 0.59 | 7.50 | 0.70 | − |
| (CH(NH$_2$)$_2$)PbI$_3$ | 0.89 | 0.81 | 16.8 | 2.56 | − |
| KPbAs | 0.98 | 210 | 6.05 | 12.7 | + |
| InAgS$_2$ | 1.53 | 0.04 | 3.38 | 1.73 | − |
| InAgSe$_2$ | 0.90 | 0.19 | 1.71 | 2.79 | + |
| Cs$_2$Hg$_3$I$_8$ | 2.55 | 0.69 | 5.03 | 0.58 | − |
| Na$_4$SnTe$_4$ | 1.63 | 1.45 | 0.05 | 0.58 | − |
| KIO$_3$ | 4.02 | 5.47 | 12.9 | 0.45 | − |
| B$_2$Pb$_4$O$_7$ | 3.42 | 0.17 | 1.00 | 0.40 | − |

**Table 1 | Bright-exciton candidates.** 28 double-Rashba materials identified by refining targets from our database search with density-functional-theory calculations. Five parameters are shown: the bandgap energy, $E_g$, the exciton Rashba energy, $\varepsilon_R$, the largest Kane energy, $E_p^{max}$ ($E_p$ depends on the polarization direction), the exciton Bohr radius, $a$, and the spin-texture helicity (STH).

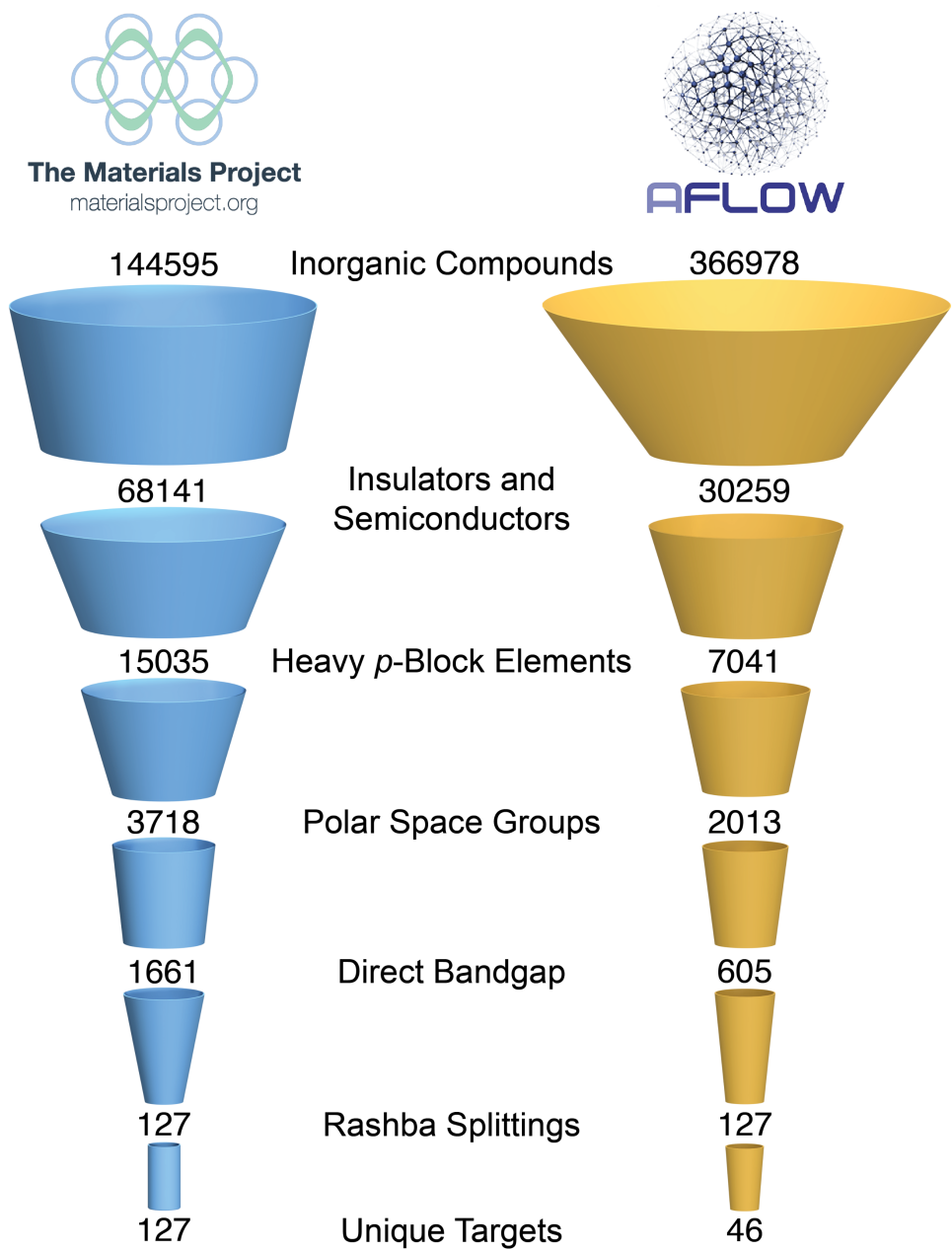

**Figure 1 | High-throughput search for 'double-Rashba' materials.** Starting with over 500,000 inorganic compounds in the Materials Project and Aflowlib databases, a series of proxy criteria for double-Rashba materials were applied moving down the diagram. After each step, the number of remaining compounds is indicated. For Aflowlib an extra step is required to eliminate duplicates. See Supplementary Section S2 and Tables S2–S4.

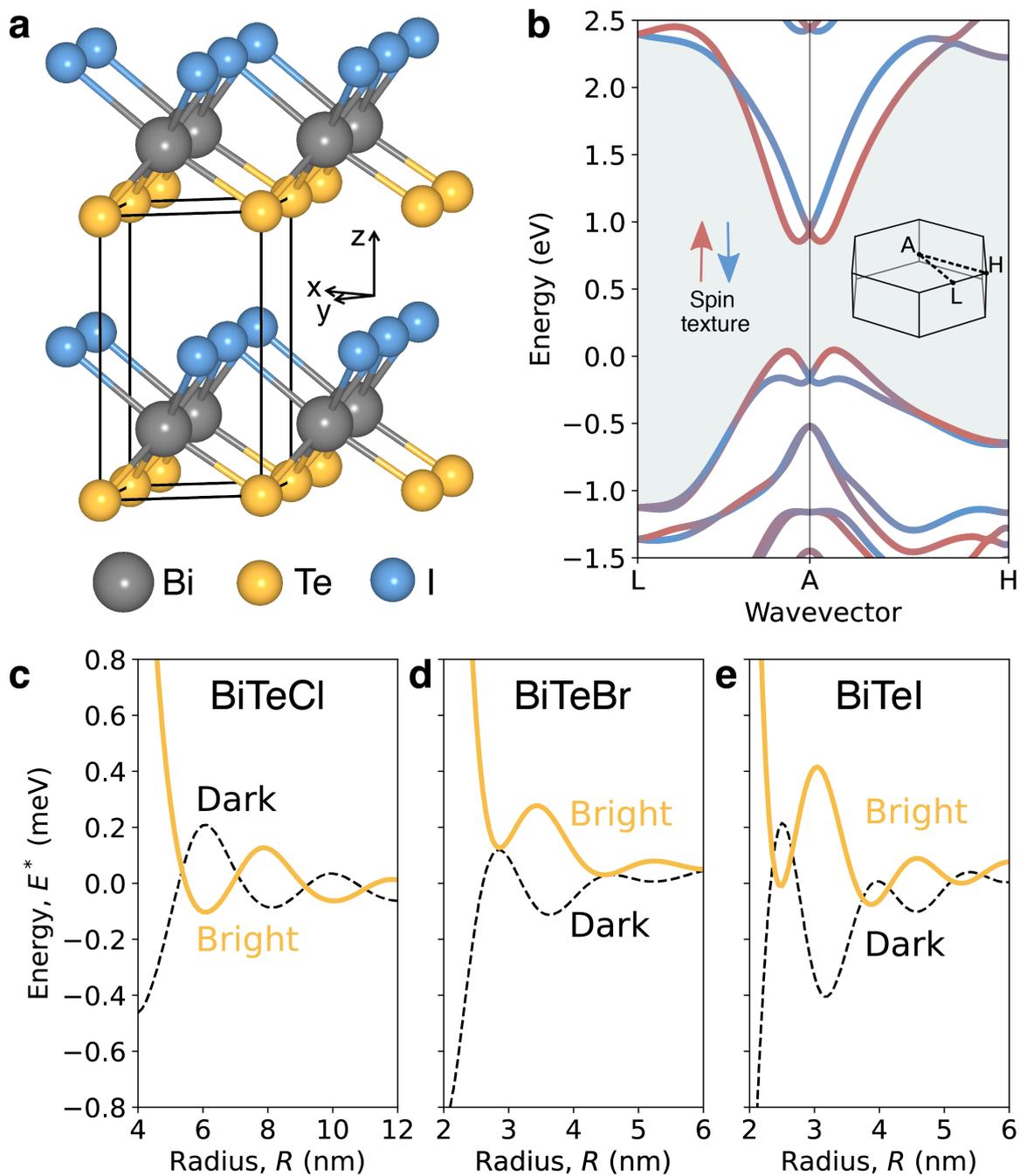

**Figure 2 | Bismuth tellurohalides (BiTeX). a**, Crystal structure of BiTeI: sheets of Te, Bi, and I form 2D monolayers of BiTeI, which then interact with neighboring layers *via* van der Waals forces. **b**, Calculated band structure of BiTeI. The inset shows the high-symmetry paths in the Brillouin zone. The central Rashba point is at the A point, and both the conduction and valence bands show Rashba splittings. The expectation value of the spin in the $y$ direction is plotted for each band (red is positive, blue is negative). **c–e**, The lowest-energy bright (yellow) and dark (dashed) exciton levels for a monolayer disk of BiTeCl, BiTeBr, and BiTeI, respectively, *versus* disk radius. The quadratic size dependence of the level energies was subtracted to focus on the oscillations. See Supplementary Section S4 for further details.

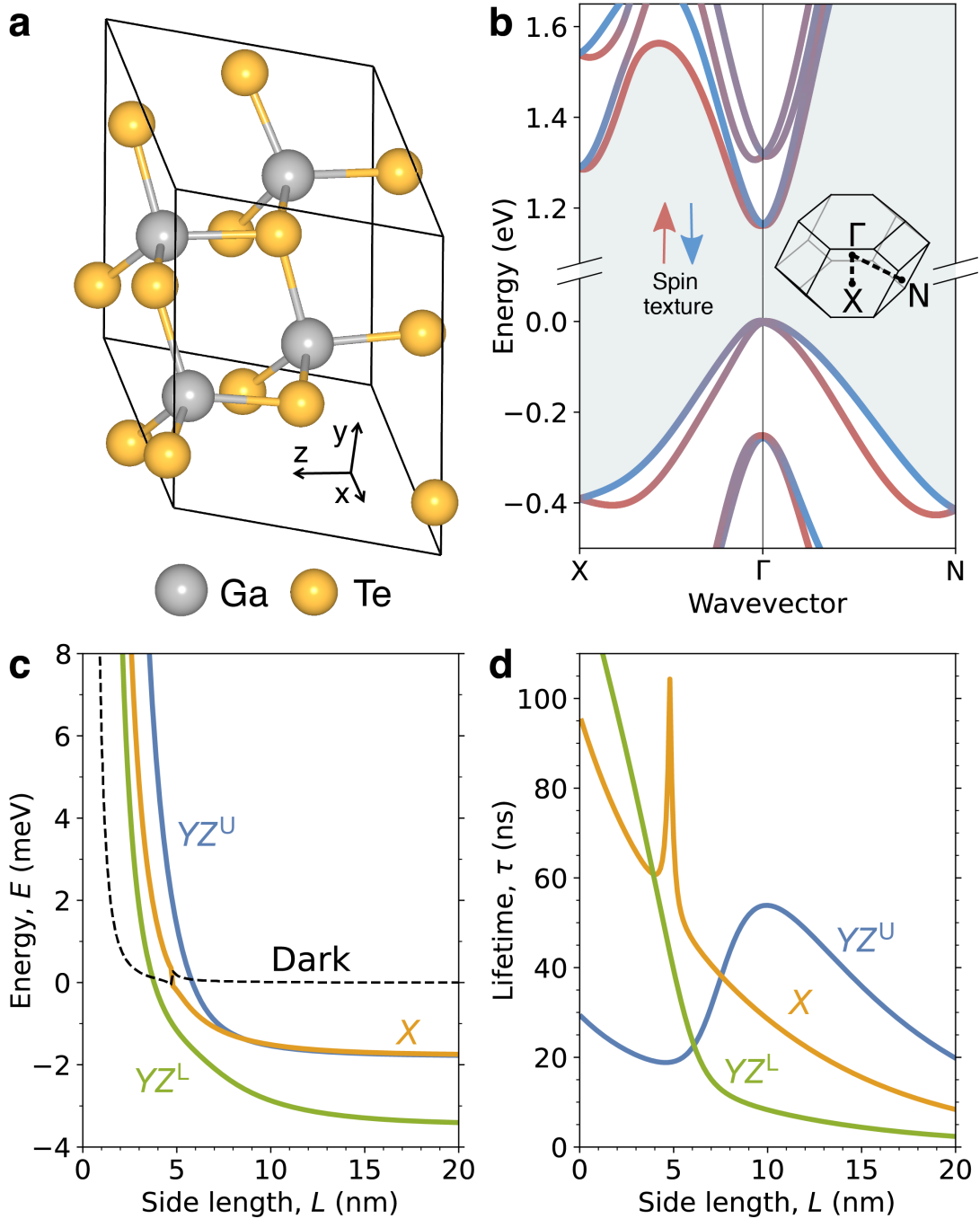

**Figure 3 | Gallium(III) telluride. a**, Crystal structure of $Ga_2Te_3$: a distorted zinc-blende lattice with ordered gallium vacancies. **b**, Calculated band structure, as in Fig. 2b, with double-Rashba behavior at Γ and the expectation value of the spin in the $y$ direction plotted for each band. **c**, Fine structure of the exciton ground state. Fine-structure energy corrections (referenced to the bulk dark state) are plotted *versus* the side length, $L$, of cube-shaped $Ga_2Te_3$ nanocrystals. The bright states (solid lines) are labeled by their polarization: $YZ^U$ and $YZ^L$ are the upper and lower states that emit light polarized in the $yz$ plane, and $X$ emits light polarized along $x$. The ground state exciton is bright for $L > 3.8$ nm. **d**, Radiative lifetime *versus* $L$ for the bright states, labeled as in **c**. See Supplementary Section S6 for further details.

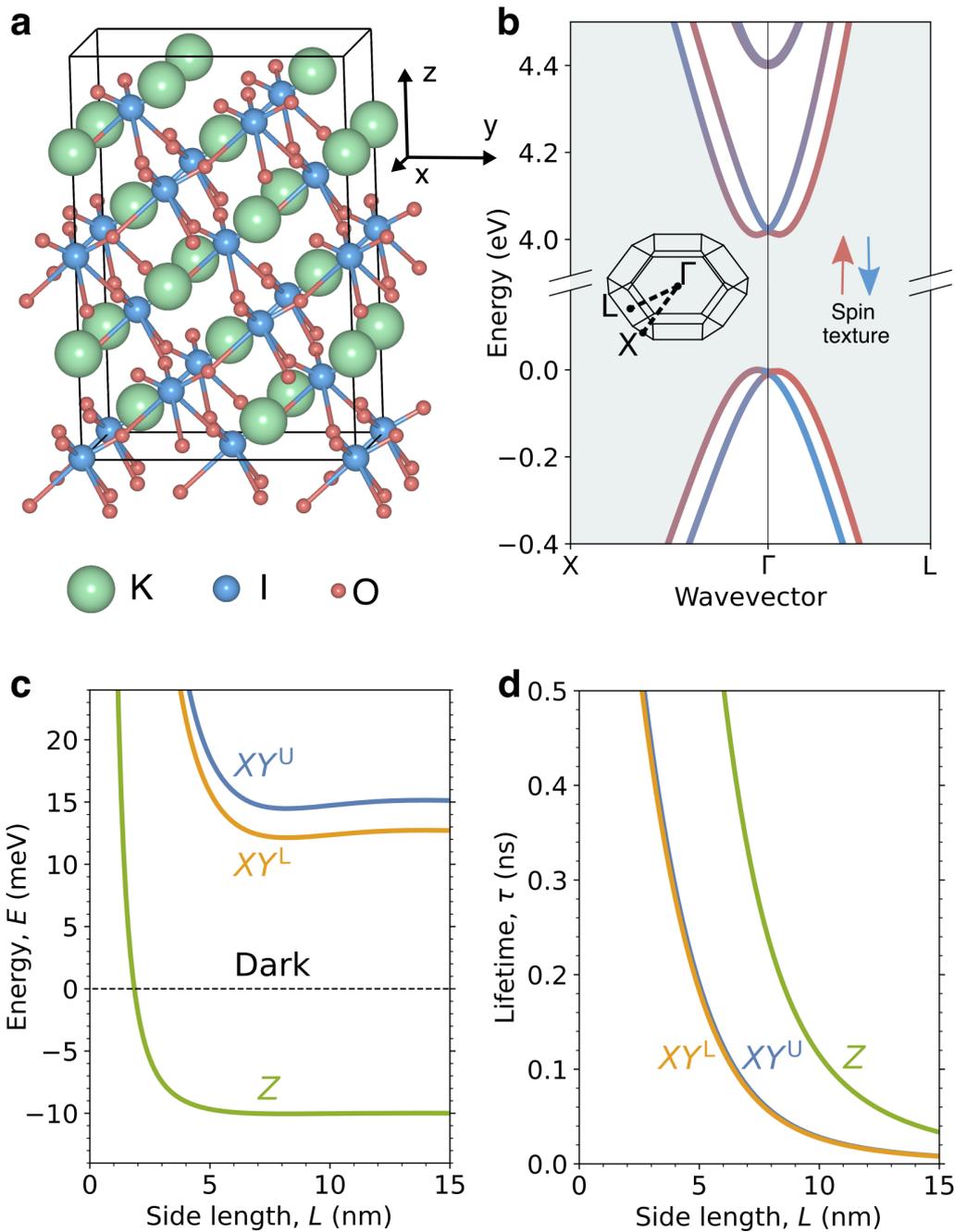

**Figure 4 | Potassium iodate. a**, Crystal structure of KIO$_3$: a distorted perovskite lattice containing corner-sharing IO$_6$ octahedra with 3 short and 3 long I–O bonds. **b**, Calculated band structure, as in Fig. 2b, with double-Rashba behavior at Γ and the expectation value of the spin in the $x$ direction plotted for each band. **c**, Fine structure of the exciton ground state. Fine-structure energy corrections (referenced to the dark state) are plotted *versus* side length, $L$, of cube-shaped KIO$_3$ nanocrystals. The bright states (solid lines) are labeled by their polarization: $XY^U$ and $XY^L$ are the upper and lower states that emit light polarized in the $xy$ plane, and $Z$ emits $z$-polarized light. **d**, Radiative lifetime *versus* $L$ for the bright states, labeled as in **c**. See Supplementary Section S7 for further details.